\documentclass[
twocolumn,
]{ceurart}

\usepackage{listings}
\usepackage{CJKutf8}
\usepackage{amsmath}
\usepackage{mathrsfs}
\usepackage{graphicx}
\usepackage{float} 
\usepackage{subfigure}

\begin{document}
	\begin{CJK*}{UTF8}{gbsn}
		\copyrightyear{2022}
		\copyrightclause{Copyright for this paper by its authors.
			Use permitted under Creative Commons License Attribution 4.0
			International (CC BY 4.0).}

		\conference{IJCAI 2023 Workshop on Deepfake Audio Detection and Analysis (DADA 2023), August 19, 2023, Macao, S.A.R}

		\title{Multi-perspective Information Fusion Res2Net with Random Specmix for Fake Speech Detection}
		
		
		\author[1]{Shunbo Dong}[%
		orcid=0000-0002-0971-3272,
		email=e22201030@stu.ahu.edu.cn,
		]
		\fnmark[1]
		\address[1]{Anhui Province Key Laboratory of Multimodal Cognitive Computation, School of Computer Science and Technology, Anhui University (AHU), 11 Jiulong Road, Hefei, 230601, China}
		
		\author[1]{Jun Xue}[%
		email=e21201068@stu.ahu.edu.cn,
		]
		\fnmark[1]
		
		\author[1]{Cunhang Fan}[%
		email=cunhang.fan@ahu.edu.cn,
		]
		\cormark[1]

		\author[1]{Kang Zhu}[%
		email=e22201061@stu.ahu.edu.cn,
		]
		
		\author[1]{Yujie Chen}[%
		email=e22201148@stu.ahu.edu.cn,
		]
		\author[1]{Zhao Lv}[%
		email=kjlz@ahu.edu.cn,
		]

		\cormark[1]

		\cortext[1]{Corresponding author.}
		\fntext[1]{These authors contributed equally.}
		
		\begin{abstract}
			In this paper, we propose the multi-perspective information fusion (MPIF) Res2Net with random Specmix for fake speech detection (FSD). The main purpose of this system is to improve the model's ability to learn precise forgery information for FSD task in low-quality scenarios. The task of random Specmix, a data augmentation, is to improve the generalization ability of the model and enhance the model's ability to locate discriminative information. Specmix cuts and pastes the frequency dimension information of the spectrogram in the same batch of samples without introducing other data, which helps the model to locate the really useful information. At the same time, we randomly select samples for augmentation to reduce the impact of data augmentation directly changing all the data. Once the purpose of helping the model to locate information is achieved, it is also important to reduce unnecessary information. The role of MPIF-Res2Net is to reduce redundant interference information. Deceptive information from a single perspective is always similar, so the model learning this similar information will produce redundant spoofing clues and interfere with truly discriminative information. The proposed MPIF-Res2Net fuses information from different perspectives, making the information learned by the model more diverse, thereby reducing the redundancy caused by similar information and avoiding interference with the learning of discriminative information. The results on the ASVspoof 2021 LA dataset demonstrate the effectiveness of our proposed method, achieving EER and min-tDCF of 3.29\% and 0.2557, respectively.

		\end{abstract}
		

		\maketitle
		
		\section{Introduction}

		Automatic speaker verification (ASV) \cite{ref1} is a technology that verifies whether a person's voice matches their voiceprint model. ASV systems are currently vulnerable to three types of attack methods: audio replay, text-to-speech (TTS), and voice conversion (VC). In order to prevent these technologies from being abused and posing a threat to social security, researchers have noticed this issue and the biennial ASVspoof challenge is held to promote the development of countermeasures. The latest competition was held in 2021\cite{ref5}, and the first started in 2015 \cite{ref2, ref3, ref4}. The first audio deepfake detection \cite{ref6} challenge was successfully held in 2022.

		Existing studies \cite{ref7, ref8, ref9, ref10, ref32} aim to propose a system that can be universally applied to synthesis speech with the unknown attack types in the clean scenarios. Regarding the scenarios with poor quality, it introduces various interference to challenge the generalization of countermeasures, and the methods in the  studies mentioned above will decrease their performance under this circumstance. For the speeches in the low-quality scenarios, researchers try to use data augmentation (DA) methods to improve the robustness of fake speech detection (FSD) task. For example, Tak et al. \cite{ref11} proposed a data boosting method, Rawboost, on the raw audio for the reliable system. This technique improved performance of FSD task greatly. Park et al. \cite{ref12} simply acted the masking method on log mel spectrogram, entailed the system to maintain robustness in the face of incomplete frequency information. In \cite{ref13}, they use frequency feature masking (FFM) to cover the information of spectrogram. Kim et al. \cite{ref20} proposed the Specmix for the spectral correlation by applying time-frequency masks. However, these DA methods are always conducted for all samples, this may affect the original data distribution characteristics, resulting in performance degradation.

		Efficient classification models have always been the research subject on FSD task. Light convolution neural network (LCNN) \cite{ref14} can separate the noise signal and information signal, and is helpful for the feature selection. Alzantot et al. \cite{ref15} built three variants based on the residual convolution network for performance improvement. Lai et al. \cite{ref16} proposed the use of SE-Net to detect speech forgery, which assigns weights according to global attention dimensions. Res2Net \cite{ref17} was proposed by Gao et al. to enhance its ability to capture multi-scale information by transferring information among channel groups. This enables the model to acquire more comprehensive information of features. Li et al. \cite{ref18} investigated the effectiveness of Res2Net in conjunction with different acoustic features. Li et al.\cite{ref19} proposed a channel-wise gating mechanism to suppress channels with lower correlations which they thought not useful. However, the models mentioned above may not achieve better results in the low-quality scenarios as they only conducted their experiments in the celan scenarios.
		
		\begin{figure*}[h]
			\centering  
			\subfigure[Res2Net backbone]{
				\label{Fig.sub.1}
				\includegraphics[width=0.28\textwidth]{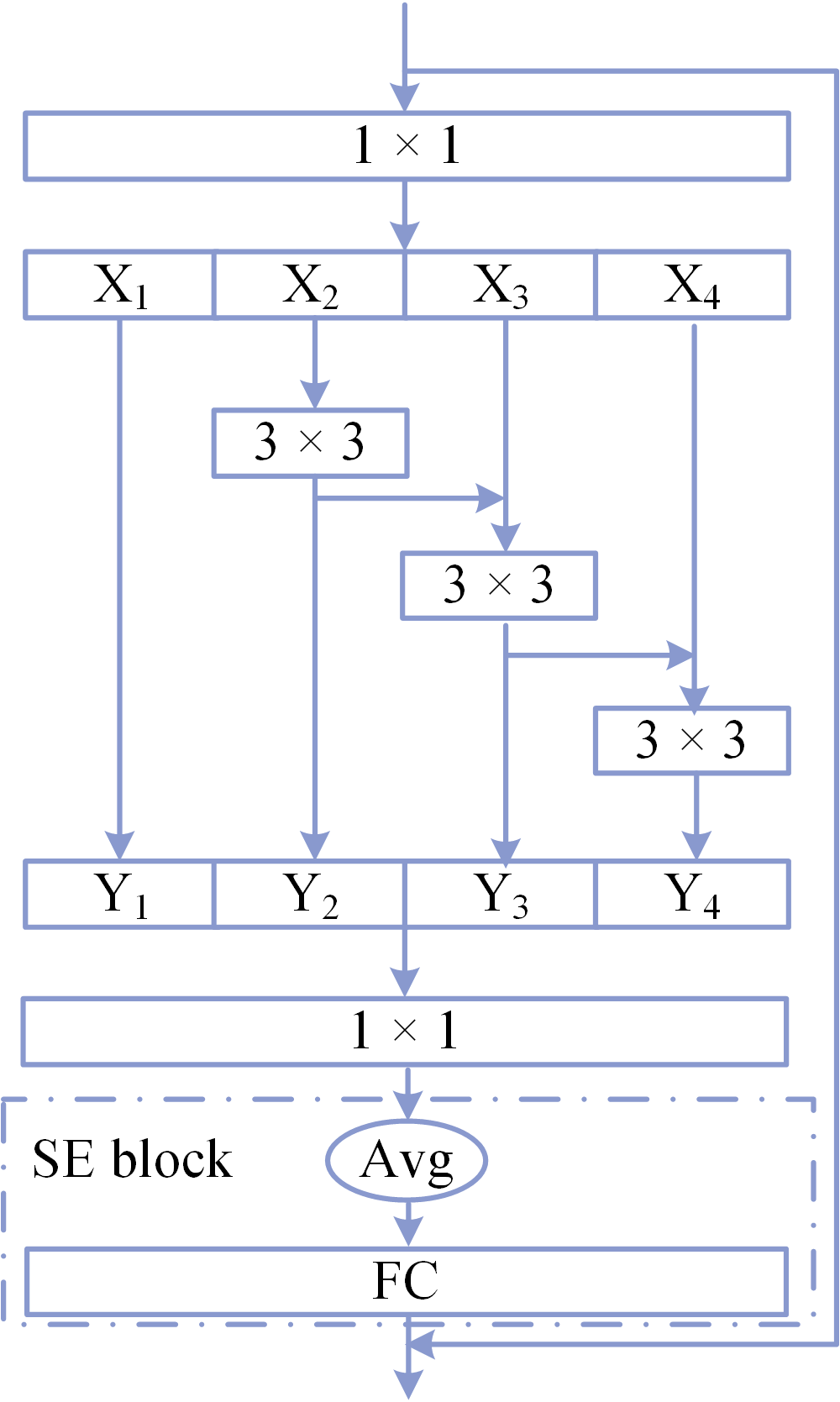}}\subfigure[MPIF-Res2Net]{
				\label{Fig.sub.2}  \hspace{8mm}
				\includegraphics[width=0.62\textwidth]{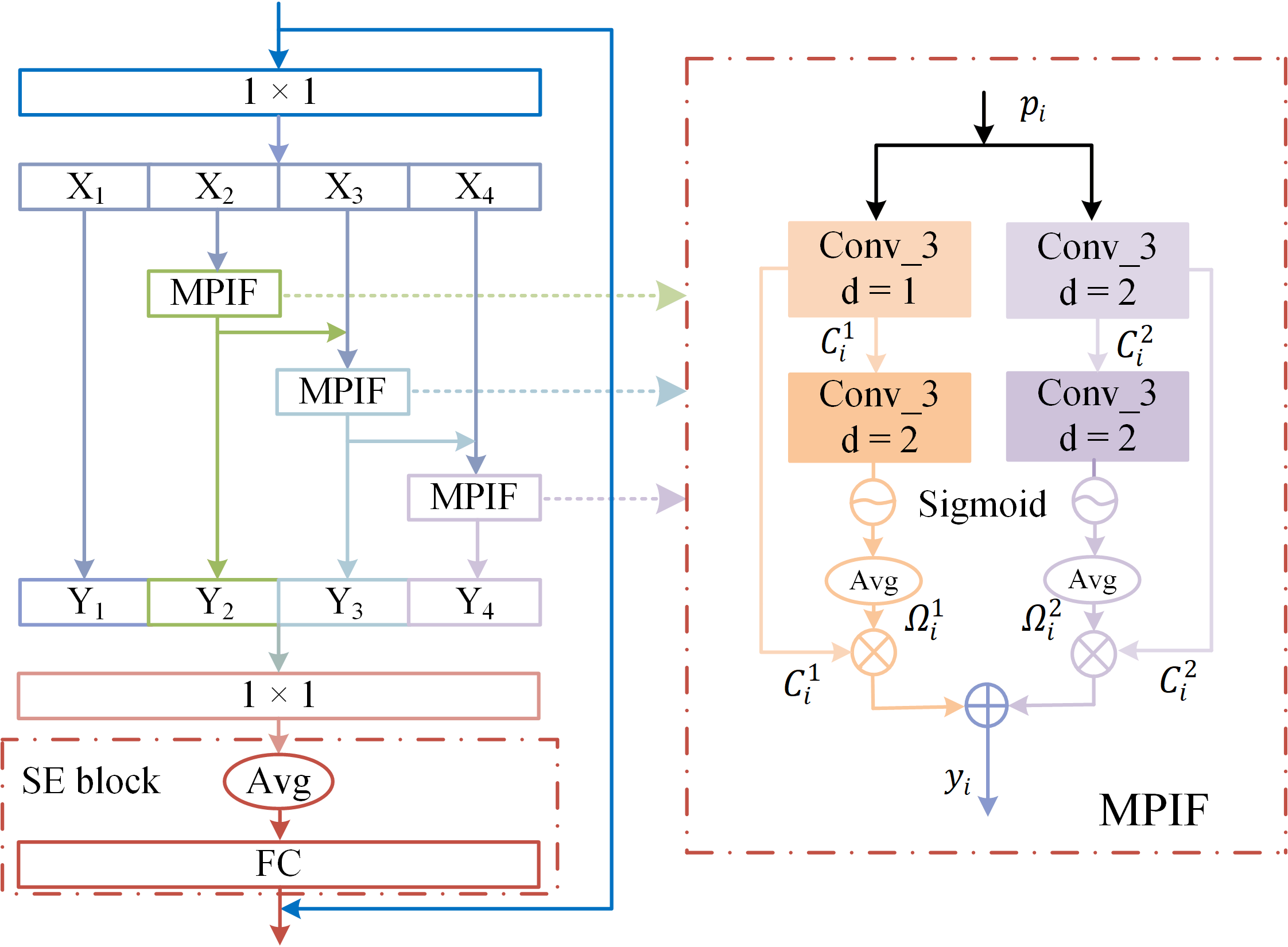}}
			\caption{Illustration of Res2Net backbone(a) and the proposed MPIF-Res2Net(b). (SE Block: the squeeze-and-excitation block \cite{ref31}; Avg is the AdaptiveAvgPool2d function; FC is full-connection layer; Sigmoid is a activate function; the MPIF-Res2Net we proposed is shown on the left of Figure ~\ref{Fig.sub.2}; the MPIF module is shown on the right of Figure ~\ref{Fig.sub.2}.)}
			\label{1}
		\end{figure*}

		In this work, we propose multi-perspective information fusion Res2Net (MPIF-Res2Net) with random Specmix. Spoofing information from a single perspective is always similar during learning process, which causes redundant information and blurs the truly discriminative information. The MPIF module fuses the information from different receptive field to reduce the redundant spoofing cues and enhance the robustness of system in the poor-quality scenarios. Specmix can increase the diversity of training data, thereby improving the generalization ability of the model. The generated spectrogram will incorporate information from other spectrograms, allowing the model to pay attention to the noteworthy information. And it performs cut and paste among spectrograms without introducing data that was not present in the original dataset with a modest impact on the original dataset. DA method conducted on all the samples may affect the distribution characteristics of the original data. For this issue, we randomly choose samples according to the probability $p\_hyper$ in advance to conduct Specmix to prevent excessive augmentation methods from weakening the fitting ability of system. Specifically, we randomly cover the part of the frequency information with corresponding frequency information of another sample in the same batch. This approach can improve the performance greatly. Our proposed method has been shown to be effective on the ASVspoof 2021 LA dataset, with achieved EER and min-tDCF results of 3.29\% and 0.2557, respectively.
		\section{Methodology}

		\subsection{Proposed method}
		In this section, we introduce the structure of the proposed MPIF-Res2Net, it reduces redundancy caused by learning single-perspective forgery information by integrating information from multiple perspectives. The convolutional operations with single kernel size are learning the similar forgery clues, producing too much redundant information and obscuring the important discriminative information. Therefore, the MPIF-Res2Net as shown on the left of Figure ~\ref{Fig.sub.2} is proposed to fuse information from different convolutional operations with different kernel size. The architecture of Res2Net is shown as Figure ~\ref{Fig.sub.1}, the outcome from the 1 $\times$ 1 convolution was splited into $n$ equal parts by the channel dimension, denoted as $p_i$, where $i$ is the integer between 1 to n. And each part has $p$ (Eq.~\ref{con:s}) channels.
		\begin{equation}
			p = \frac{\#channel}{n}\label{con:s}
		\end{equation}
		where $\#channel$ means the total number of channels.
		Res2Net uses the residual-like connection to perform addition between the channel groups. The following formulation can be used to describe this process:
		\begin{equation}
			y_{i}=\left\{\begin{array}{ll}
				p_{i}, & i=1 \\
				K_{i}\left(p_{i}\right), & i=2 \\
				K_{i}\left(p_{i}+y_{i-1}\right), & 2<i \leq n
			\end{array}\right.
		\end{equation}
		Where $K_i$ represents the convolution operation. The proposed MPIF-Res2Net replaces the $K_i$ operation with MPIF module as shown on the right of Figure ~\ref{Fig.sub.2}. The $y_i$ is calculated as fowllows:
		\begin{equation}
			y_{i}=\left\{\begin{array}{ll}
				p_{i}, & i=1 \\
				MPIF_{i}\left(p_{i}\right), & i=2 \\
				MPIF_{i}\left(p_{i}+y_{i-1}\right), & 2<i \leq n
			\end{array}\right.
		\end{equation}
		
		\subsection{Multi-perspective Information Fusion  Module}
		As shown on the right of Figure ~\ref{Fig.sub.2}, MPIF module in current channel group $i$ performs different convolution operation on $p_i$ or $p_i+y_{i-1}$ to get the spoofing information from different perspective. Firstly, $p_i$ is sent into the  convolution operations with different dilation parameter $j$, where $j\in[1,2]$, at the beginning of MPIF module (Eq.~\ref{con:4}). The results are then passing through the dilated convolution to recalculate the energy distribution of each channel, normalize them through the Sigmoid function, and then the average pooling layer is used to get the results $\omega_k^j$ as the weight of each channel $k$ from $Conv2d$. And the purpose of using dilated convolution is to increase the receptive field, ensuring that each convolution output contains information from a larger range while keeping the parameter and computation cost constant. The weighting factor $\omega_k^j$ of each channel $k$ is calculated by Eq.~\ref{con:5}.
		\begin{equation}    
			c^j = \mathcal (Conv2d_j (p_i))) \label{con:4}
		\end{equation}
		
		\begin{equation}
			\omega_k^j = \mathcal Avg (Sigmoid (Conv2d (c_k^j))) \label{con:5}
		\end{equation}
		$Conv2d_j$ at the beginning of MPIF module takes $p_i$ as input and outputs $c^j$. $c_k^j$ is the $k\-th$ channel in $conv^j$. $Conv2d$ is a convolutional operation to recalculate energy distribution. $Sigmoid$ is the $Sigmoid$ function. $Avg$ denotes the $AdaptiveAvgPool2d$ function.

		After the weighting factor $\omega_k^j$, we perform multiplication on $c_k^j$ and $\omega_k^j$, and sum up the results. Then we can get the result $MPIF_i(p_i)$ of $i$-th channel group as follows:
		\begin{equation}
			{MPIF_i(p_i)}=\sum_{j=1}^2{C_i^j} \times ({\Omega_i^j })
		\end{equation}
		where $C_i^j$ is a matrix composed with $c_k^j$, 
		$\Omega_i^j$ $\in \mathbb{R}^{p \times 1 \times 1}$ is the weight matrix with $\omega_k^j$. The $p$ is the number of the channel of $p_i$.

		\begin{table}
			\caption{The Proposed MPIF-Res2net Model Architecture and
				Configuration. the Dimensions Are Arranged in the Order of Channels, Frequency, and Time). BN Denotes Batch Normalization and ReLU denotes Rectified Linear Unit, MPIF and SE Are the Multi-perspective Information Module and the Squeeze And Excitation Layer, Respectively.}
			
			\label{tab:network}
			\resizebox{0.49\textwidth}{46.1mm}{
				\begin{tabular}{cllllclllllcllll}
					\hline
					\multicolumn{5}{c}{Layer}                                                                       & \multicolumn{6}{c}{Input:27000 samples}                                                                                 & \multicolumn{5}{c}{Output shape}                                                                                      \\ \hline
					\multicolumn{5}{c}{Front-end}                                                                   & \multicolumn{6}{c}{F0 subband}                                                                                          & \multicolumn{5}{c}{(45,600)(F,T)}                                                                                     \\ \hline
					\multicolumn{5}{c}{\multirow{3}{*}{Pre-processing}}                                             & \multicolumn{6}{c}{\multirow{3}{*}{\begin{tabular}[c]{@{}c@{}}Channel expansion\\ Conv2D\_1\\ BN \& ReLU\end{tabular}}} & \multicolumn{5}{c}{\multirow{3}{*}{\begin{tabular}[c]{@{}c@{}}(1,45,600)\\ (16,45,600)\end{tabular}}}                 \\
					\multicolumn{5}{c}{}                                                                            & \multicolumn{6}{c}{}                                                                                                    & \multicolumn{5}{c}{}                                                                                                  \\
					\multicolumn{5}{c}{}                                                                            & \multicolumn{6}{c}{}                                                                                                    & \multicolumn{5}{c}{}                                                                                                  \\ \hline
					\multicolumn{5}{c}{\multirow{10}{*}{\begin{tabular}[c]{@{}c@{}}Layer1\\ \&Layer3\end{tabular}}} & \multicolumn{6}{c}{\multirow{5}{*}{\begin{tabular}[c]{@{}c@{}}1 $\times$ $\begin{cases}Conv2D\_1\\ Conv2D\_3\\ Conv2D\_1\\ SE\end{cases}$\end{tabular}}}     & \multicolumn{5}{c}{\multirow{10}{*}{\begin{tabular}[c]{@{}c@{}}Layer1 (32,45,600)\\ Layer3(128,12,150)\end{tabular}}} \\
					\multicolumn{5}{c}{}                                                                            & \multicolumn{6}{c}{}                                                                                                    & \multicolumn{5}{c}{}                                                                                                  \\
					\multicolumn{5}{c}{}                                                                            & \multicolumn{6}{c}{}                                                                                                    & \multicolumn{5}{c}{}                                                                                                  \\
					\multicolumn{5}{c}{}                                                                            & \multicolumn{6}{c}{}                                                                                                    & \multicolumn{5}{c}{}                                                                                                  \\
					\multicolumn{5}{c}{}                                                                            & \multicolumn{6}{c}{}                                                                                                    & \multicolumn{5}{c}{}                                                                                                  \\ 
					\multicolumn{5}{c}{}                                                                            & \multicolumn{6}{c}{\multirow{5}{*}{\begin{tabular}[c]{@{}c@{}}1 $\times$$\begin{cases}Conv2D\_1\\ MPIF\\ Conv2D\_1\\ SE\end{cases}$\end{tabular}}}          & \multicolumn{5}{c}{}                                                                                                  \\
					\multicolumn{5}{c}{}                                                                            & \multicolumn{6}{c}{}                                                                                                    & \multicolumn{5}{c}{}                                                                                                  \\
					\multicolumn{5}{c}{}                                                                            & \multicolumn{6}{c}{}                                                                                                    & \multicolumn{5}{c}{}                                                                                                  \\
					\multicolumn{5}{c}{}                                                                            & \multicolumn{6}{c}{}                                                                                                    & \multicolumn{5}{c}{}                                                                                                  \\
					\multicolumn{5}{c}{}                                                                            & \multicolumn{6}{c}{}                                                                                                    & \multicolumn{5}{c}{}                                                                                                  \\ \hline
					\multicolumn{5}{c}{\multirow{10}{*}{\begin{tabular}[c]{@{}c@{}}Layer2\\ \&Layer4\end{tabular}}} & \multicolumn{6}{c}{\multirow{5}{*}{\begin{tabular}[c]{@{}c@{}}1 $\times$$\begin{cases}Conv2D\_1\\ Conv2D\_3\\ Conv2D\_1\\ SE\end{cases}$\end{tabular}}}     & \multicolumn{5}{c}{\multirow{10}{*}{\begin{tabular}[c]{@{}c@{}}Layer2(64,23,300)\\ Layer4(256,6,75)\end{tabular}}}    \\
					\multicolumn{5}{c}{}                                                                            & \multicolumn{6}{c}{}                                                                                                    & \multicolumn{5}{c}{}                                                                                                  \\
					\multicolumn{5}{c}{}                                                                            & \multicolumn{6}{c}{}                                                                                                    & \multicolumn{5}{c}{}                                                                                                  \\
					\multicolumn{5}{c}{}                                                                            & \multicolumn{6}{c}{}                                                                                                    & \multicolumn{5}{c}{}                                                                                                  \\
					\multicolumn{5}{c}{}                                                                            & \multicolumn{6}{c}{}                                                                                                    & \multicolumn{5}{c}{}                                                                                                  \\ 
					\multicolumn{5}{c}{}                                                                            & \multicolumn{6}{c}{\multirow{5}{*}{\begin{tabular}[c]{@{}c@{}}2 $\times\begin{cases}Conv2D\_1\\ MPIF\\ Conv2D\_1\\ SE\end{cases}$\end{tabular}}}          & \multicolumn{5}{c}{}                                                                                                  \\
					\multicolumn{5}{c}{}                                                                            & \multicolumn{6}{c}{}                                                                                                    & \multicolumn{5}{c}{}                                                                                                  \\
					\multicolumn{5}{c}{}                                                                            & \multicolumn{6}{c}{}                                                                                                    & \multicolumn{5}{c}{}                                                                                                  \\
					\multicolumn{5}{c}{}                                                                            & \multicolumn{6}{c}{}                                                                                                    & \multicolumn{5}{c}{}                                                                                                  \\
					\multicolumn{5}{c}{}                                                                            & \multicolumn{6}{c}{}                                                                                                    & \multicolumn{5}{c}{}                                                                                                  \\ \hline
					\multicolumn{5}{c}{\multirow{2}{*}{Output}}                                                     & \multicolumn{6}{c}{\multirow{2}{*}{\begin{tabular}[c]{@{}c@{}}Avgpool2D(1, 1)\\ AngleLinear\end{tabular}}}               & \multicolumn{5}{c}{\multirow{2}{*}{\begin{tabular}[c]{@{}c@{}}(256,1,1)\\ 2\end{tabular}}}                            \\
					\multicolumn{5}{c}{}                                                                            & \multicolumn{6}{c}{}                                                                                                    & \multicolumn{5}{c}{}                                                                                                  \\ \hline
			\end{tabular}}
		\end{table}

		\subsection{Random Specmix Strategy}
		In this work, we use a random Specmix strategy to help the model to locate the discriminative information and enhance the generalization of the model. For the training of deep neural networks, we always transform the raw audio from time domain into time-frequency domain. And inspired by \cite{ref20}, we conduct Specmix on the frequency dimension of the F0 subband \cite{ref30}, which is a subband of amplitude spectrum, and the maximum span of Specmix operation is no more than 10. Specmix cuts and pastes spectrograms among themselves in the same batch to help the model focus on the discriminative regions that may be worth to attend to. And different from \cite{ref20}, there is no Specmix operation on labels. We cover the information on frequency dimension with the corresponding parts of other samples in the same batch. At the same time, to avoid the conduction of Specmix on all the samples, inspired by \cite{ref21}, we randomly choose speech samples according to the hyperparameter $p\_hyper$ in advance to conduct Specmix operation. For a batch of speech samples, the probability of them undergoing random Specmix is $p$, when $p$ is bigger than $p\_hyper$, Specmix was conducted on them, otherwise no conduction with Specmix. And in the evaluation phase, we do not use the random Specmix strategy.
		
		\begin{table*}[h]
			\caption{Results of Ablation Experiments for Our Proposal Module. $(p\_hyper)$ Means the Probability of Application of random Specmix strategy. Res2Net\_k3 Denotes the Kernel Sizes of Convolution Are 3 In the channel groups; Res2Net\_k5 Denotes the Kernel Sizes of Convolution Are 3 with the Dilation Parameter Is 2 In the channel groups. }
			\label{tab:result}
			\resizebox{0.8\textwidth}{48mm}{
				\begin{tabular}{ccccccc}
					\hline
					\multicolumn{7}{c}{\multirow{2}{*}{Ablation Experiments Results On ASVspoof 2021 LA dataset}}                                                                                                                                                                                                                                       \\
					\multicolumn{7}{c}{}                               \\ \hline
					\multicolumn{1}{c|}{\multirow{2}{*}{$(p\_hyper)$}} & \multicolumn{2}{c|}{MPIF-Res2Net}                                                                           & \multicolumn{2}{c|}{Res2Net\_k3}                                                          & \multicolumn{2}{c}{Res2Net\_k5}                                      \\ \cline{2-7} 
					\multicolumn{1}{c|}{}                              & \multicolumn{1}{c|}{EER(\%)}                        & \multicolumn{1}{c|}{min-tDCF}                         & \multicolumn{1}{c|}{EER(\%)}               & \multicolumn{1}{c|}{min-tDCF}                & \multicolumn{1}{c|}{EER(\%)}               & min-tDCF                \\ \hline
					\multicolumn{1}{c|}{\multirow{2}{*}{0}}            & \multicolumn{1}{c|}{\multirow{2}{*}{4.04}}          & \multicolumn{1}{c|}{\multirow{2}{*}{0.2713}}          & \multicolumn{1}{c|}{\multirow{2}{*}{4.26}} & \multicolumn{1}{c|}{\multirow{2}{*}{0.2750}} & \multicolumn{1}{c|}{\multirow{2}{*}{4.00}} & \multirow{2}{*}{0.2702} \\
					\multicolumn{1}{c|}{}                              & \multicolumn{1}{c|}{}                               & \multicolumn{1}{c|}{}                                 & \multicolumn{1}{c|}{}                      & \multicolumn{1}{c|}{}                        & \multicolumn{1}{c|}{}                      &                         \\ \hline
					\multicolumn{1}{c|}{\multirow{2}{*}{0.1}}          & \multicolumn{1}{c|}{\multirow{2}{*}{3.57}}          & \multicolumn{1}{c|}{\multirow{2}{*}{0.2577}}          & \multicolumn{1}{c|}{\multirow{2}{*}{4.16}} & \multicolumn{1}{c|}{\multirow{2}{*}{0.2760}} & \multicolumn{1}{c|}{\multirow{2}{*}{4.44}} & \multirow{2}{*}{0.2806} \\
					\multicolumn{1}{c|}{}                              & \multicolumn{1}{c|}{}                               & \multicolumn{1}{c|}{}                                 & \multicolumn{1}{c|}{}                      & \multicolumn{1}{c|}{}                        & \multicolumn{1}{c|}{}                      &                         \\ \hline
					\multicolumn{1}{c|}{\multirow{2}{*}{0.2}}          & \multicolumn{1}{c|}{\multirow{2}{*}{3.96}}          & \multicolumn{1}{c|}{\multirow{2}{*}{0.2693}}          & \multicolumn{1}{c|}{\multirow{2}{*}{4.29}} & \multicolumn{1}{c|}{\multirow{2}{*}{0.2739}} & \multicolumn{1}{c|}{\multirow{2}{*}{4.08}} & \multirow{2}{*}{0.2711} \\
					\multicolumn{1}{c|}{}                              & \multicolumn{1}{c|}{}                               & \multicolumn{1}{c|}{}                                 & \multicolumn{1}{c|}{}                      & \multicolumn{1}{c|}{}                        & \multicolumn{1}{c|}{}                      &                         \\ \hline
					\multicolumn{1}{c|}{\multirow{2}{*}{0.3}}          & \multicolumn{1}{c|}{\multirow{2}{*}{3.98}}          & \multicolumn{1}{c|}{\multirow{2}{*}{0.2692}}          & \multicolumn{1}{c|}{\multirow{2}{*}{4.08}} & \multicolumn{1}{c|}{\multirow{2}{*}{0.2753}} & \multicolumn{1}{c|}{\multirow{2}{*}{4.18}} & \multirow{2}{*}{0.2762} \\
					\multicolumn{1}{c|}{}                              & \multicolumn{1}{c|}{}                               & \multicolumn{1}{c|}{}                                 & \multicolumn{1}{c|}{}                      & \multicolumn{1}{c|}{}                        & \multicolumn{1}{c|}{}                      &                         \\ \hline
					\multicolumn{1}{c|}{\multirow{2}{*}{0.4}}          & \multicolumn{1}{c|}{\multirow{2}{*}{3.70}}           & \multicolumn{1}{c|}{\multirow{2}{*}{0.2638}}          & \multicolumn{1}{c|}{\multirow{2}{*}{3.87}} & \multicolumn{1}{c|}{\multirow{2}{*}{0.2688}} & \multicolumn{1}{c|}{\multirow{2}{*}{4.08}} & \multirow{2}{*}{0.2719} \\
					\multicolumn{1}{c|}{}                              & \multicolumn{1}{c|}{}                               & \multicolumn{1}{c|}{}                                 & \multicolumn{1}{c|}{}                      & \multicolumn{1}{c|}{}                        & \multicolumn{1}{c|}{}                      &                         \\ \hline
					\multicolumn{1}{c|}{\multirow{2}{*}{0.5}}          & \multicolumn{1}{c|}{\multirow{2}{*}{\textbf{3.29}}} & \multicolumn{1}{c|}{\multirow{2}{*}{\textbf{0.2557}}} & \multicolumn{1}{c|}{\multirow{2}{*}{4.23}} & \multicolumn{1}{c|}{\multirow{2}{*}{0.2738}} & \multicolumn{1}{c|}{\multirow{2}{*}{4.24}} & \multirow{2}{*}{0.2756} \\
					\multicolumn{1}{c|}{}                              & \multicolumn{1}{c|}{}                               & \multicolumn{1}{c|}{}                                 & \multicolumn{1}{c|}{}                      & \multicolumn{1}{c|}{}                        & \multicolumn{1}{c|}{}                      &                         \\ \hline
					\multicolumn{1}{c|}{\multirow{2}{*}{0.6}}          & \multicolumn{1}{c|}{\multirow{2}{*}{3.65}}          & \multicolumn{1}{c|}{\multirow{2}{*}{0.2679}}          & \multicolumn{1}{c|}{\multirow{2}{*}{3.98}} & \multicolumn{1}{c|}{\multirow{2}{*}{0.2708}} & \multicolumn{1}{c|}{\multirow{2}{*}{3.54}} & \multirow{2}{*}{0.2598} \\
					\multicolumn{1}{c|}{}                              & \multicolumn{1}{c|}{}                               & \multicolumn{1}{c|}{}                                 & \multicolumn{1}{c|}{}                      & \multicolumn{1}{c|}{}                        & \multicolumn{1}{c|}{}                      &                         \\ \hline
					\multicolumn{1}{c|}{\multirow{2}{*}{0.7}}          & \multicolumn{1}{c|}{\multirow{2}{*}{3.69}}          & \multicolumn{1}{c|}{\multirow{2}{*}{0.2737}}          & \multicolumn{1}{c|}{\multirow{2}{*}{4.25}} & \multicolumn{1}{c|}{\multirow{2}{*}{0.2775}} & \multicolumn{1}{c|}{\multirow{2}{*}{3.56}} & \multirow{2}{*}{0.2660} \\
					\multicolumn{1}{c|}{}                              & \multicolumn{1}{c|}{}                               & \multicolumn{1}{c|}{}                                 & \multicolumn{1}{c|}{}                      & \multicolumn{1}{c|}{}                        & \multicolumn{1}{c|}{}                      &                         \\ \hline
					\multicolumn{1}{c|}{\multirow{2}{*}{0.8}}          & \multicolumn{1}{c|}{\multirow{2}{*}{3.73}}          & \multicolumn{1}{c|}{\multirow{2}{*}{0.2692}}          & \multicolumn{1}{c|}{\multirow{2}{*}{4.25}} & \multicolumn{1}{c|}{\multirow{2}{*}{0.2804}} & \multicolumn{1}{c|}{\multirow{2}{*}{3.58}} & \multirow{2}{*}{0.2670} \\
					\multicolumn{1}{c|}{}                              & \multicolumn{1}{c|}{}                               & \multicolumn{1}{c|}{}                                 & \multicolumn{1}{c|}{}                      & \multicolumn{1}{c|}{}                        & \multicolumn{1}{c|}{}                      &                         \\ \hline
					\multicolumn{1}{c|}{\multirow{2}{*}{0.9}}          & \multicolumn{1}{c|}{\multirow{2}{*}{4.02}}          & \multicolumn{1}{c|}{\multirow{2}{*}{0.2696}}          & \multicolumn{1}{c|}{\multirow{2}{*}{4.23}} & \multicolumn{1}{c|}{\multirow{2}{*}{0.2758}} & \multicolumn{1}{c|}{\multirow{2}{*}{4.13}} & \multirow{2}{*}{0.2731} \\
					\multicolumn{1}{c|}{}                              & \multicolumn{1}{c|}{}                               & \multicolumn{1}{c|}{}                                 & \multicolumn{1}{c|}{}                      & \multicolumn{1}{c|}{}                        & \multicolumn{1}{c|}{}                      &                         \\ \hline
					\multicolumn{1}{c|}{\multirow{2}{*}{1(no Specmix)}}   & \multicolumn{1}{c|}{\multirow{2}{*}{3.70}}          & \multicolumn{1}{c|}{\multirow{2}{*}{0.2707}}          & \multicolumn{1}{c|}{\multirow{2}{*}{4.58}} & \multicolumn{1}{c|}{\multirow{2}{*}{0.2811}} & \multicolumn{1}{c|}{\multirow{2}{*}{4.49}} & \multirow{2}{*}{0.2859} \\
					\multicolumn{1}{c|}{}                              & \multicolumn{1}{c|}{}                               & \multicolumn{1}{c|}{}                                 & \multicolumn{1}{c|}{}                      & \multicolumn{1}{c|}{}                        & \multicolumn{1}{c|}{}        &                        
					\\ \hline
			\end{tabular}}
		\end{table*}

		\section{Expriments And Results}

		\subsection{Experimental Setup}
		It must be a challenging task to learn a robust countermeasure suitable to low-quality scenario trained  on the training set without same interference conditions. In this work, we use the Rawboost \cite{ref11} DA method to train the model, this technique can enhance the accuracy in the low-quality scenarios. To be more precise, the impulsive signal-dependent (ISD) additive noise and stationary signal-independent (SSI) additive noise are added to the raw waveform. After the STFT operation with the window length is 1728 and the hop length is 130, we got a spectrogram of size 865. We then truncate or concatenate the spectrogram to fix the number of frames at 600. We utilize the 0-400 Hz LPS feature with the first 0-45 dimension as our F0 subband feature. The resulting feature size of F0 subband is 45$\times$600. Then, we determine whether to conduct random Specmix on the samples in the current batch by setting the hyperparameter $(p\_hyper)$, the probability whether to conduct the random Specmix strategy. Considering that the F0 feature is a subband of the amplitude spectrum, we set the maximum span for Specmix to be no more than 10.

		In this article, we propose MPIF-Res2Net to fuse the information from different perspective to reduce the redundant spoofing cues and introduce random Specmix to improve the generalization ability of the model. Table ~\ref{tab:network} presents the design of MPIF-Res2Net, which includes details on channels, convolution kernels, and repetition frequency. In our experiments, Adam is utilized as the optimizer, with the following parameter settings: $\beta_1$ $=$ 0.9, $\beta_2$ $=$ 0.98, $\epsilon$ $=$ $10^{-9}$, and weight decay is $10^{-4}$. The epoch is set to 32. And the number of channel groups is set to 8. The batch size is 16.
		
		\subsection{Dataset}
		The data in the ASVspoof 2019 logical access (LA) dataset is divided into three subsets: training set, development set, and evaluation set. The spoof speech in the training and development sets comes from six speech synthesis and speech conversion technologies, which are known attack types. The evaluation set contains audio generated by 11 unknown attack types. We trained our model on the ASVspoof 2019 training set and selected the best performing model on  the development set. 
		As stated in \cite{ref33}, the ASVspoof 2021 LA dataset is designed for developing anti-spoofing methods that can effectively adapt to unknown channel variations and does not provide new matching training or development data. The speech samples from the ASVspoof 2021 evaluation set were transmitted via actual telephone systems utilizing various bandwidths and codecs. The data in the 2019 LA training and development subset does not have similar encoding and transmission, and these subsets only contain clean data. Equal error rate (EER) and minimum tandem detection cost function (min t-DCF) are used as the metrics.

		\subsection{Experimental Results}
		\subsubsection{Ablation Study}

		Firstly, different values of the probability $(p\_hyper)$ should be considered as the guidance of the experimental conduction to obtain the best $p\_hyper$. Table ~\ref{tab:result} shows the EER results of conduction with random Specmix strategy for different values of $p\_hyper$. The MPIF-Res2Net with $p\_hyper$=0.5 has the best performance whose EER result is 3.29\%, and the min t-DCF result is 0.2557, which means a relatively higher reliability of the countermeasure system when it is applied with an ASV system. For experiments involving information from a single receptive field, we set up two models, Res2Net\_k3 and Res2Net\_k5, with the parameter kernel sizes and dilations are 3 and 1, 3 and 2, respectively. The EER result of MPIF-Res2Net with the $p\_hyper$ equal to 1 is 3.70\%, however, the corresponding EER results of the other two systems are 4.58\% and 4.49\% respectively, which verifies the MPIF-Res2Net we proposed do have the ability by fusing information from different perspective to reduce the redundancy caused by learning the similar spoofing clues with the single kernel size. The EER results of Res2Net\_k3 and Res2Net\_k5 undergoing Specmix demonstrate that Specmix can help help the model to locate the forgery information and improve model generalization performance.

		For random Specmix strategy, the MPIF-Res2Net with $p\_hyper$ is 0 got the EER result of 4.04\%, the $p\_hyper$ with 0 means the Specmix conduction was conducted on all of the samples, this indicates that all the samples undergoing Specmix cause the serious performance degradation of system. Overall, the random Specmix has improved the model's generalization ability and enhanced its performance.
		
		Experimental results show that our proposed MPIF-Res2Net with random Specmix enhancement methods can improve performance for FSD task in the low-quality scenarios.
		\begin{table}[]
			\caption{Results Comparison with Fusion Systems on the Performance of ASVspoof2021 Dataset}
			\label{result2}
			\begin{tabular}{ccllcll}
				\hline
				System       & \multicolumn{3}{c}{t-DCF}  & \multicolumn{3}{c}{EER(\%)} \\ \hline
				
				T23 \cite{ref22}  & \multicolumn{3}{c}{0.2177} & \multicolumn{3}{c}{1.32}    \\ \hline
				T20 \cite{ref23}  & \multicolumn{3}{c}{0.2608} & \multicolumn{3}{c}{3.21}    \\ \hline
				T04 \cite{ref24}  & \multicolumn{3}{c}{0.2747} & \multicolumn{3}{c}{5.58}    \\ \hline
				T06 \cite{ref25}  & \multicolumn{3}{c}{0.2853} & \multicolumn{3}{c}{5.66}    \\ \hline
				T35 \cite{ref22}  & \multicolumn{3}{c}{0.2480} & \multicolumn{3}{c}{2.77}    \\ \hline
				T19 \cite{ref22}  & \multicolumn{3}{c}{0.2495} & \multicolumn{3}{c}{3.13}    \\ \hline
				Fusion systems \cite{ref27}  & \multicolumn{3}{c}{0.2882} & \multicolumn{3}{c}{4.66}    \\ \hline
				MPIF-Res2Net \textbf{ours}   & \multicolumn{3}{c}{\textbf{0.2557}} & \multicolumn{3}{c}{\textbf{3.29}}    \\ \hline
			\end{tabular}
		\end{table}
		
		\begin{table}[]
			\caption{Results Comparison with Single System on the Performance of ASVspoof2021 Dataset}
			\label{result1}
			\begin{tabular}{ccllcll}
				\hline
				System       & \multicolumn{3}{c}{t-DCF}  & \multicolumn{3}{c}{EER(\%)} \\ \hline
				B03 \cite{ref28} & \multicolumn{3}{c}{0.3445} & \multicolumn{3}{c}{9.26}    \\ \hline
				B04 \cite{ref28} & \multicolumn{3}{c}{0.4257} & \multicolumn{3}{c}{9.5}     \\ \hline
				B01 \cite{ref28} & \multicolumn{3}{c}{0.4974} & \multicolumn{3}{c}{15.62}   \\ \hline
				B02 \cite{ref28} & \multicolumn{3}{c}{0.5758} & \multicolumn{3}{c}{19.3}    \\ \hline
				RawNet2 \cite{ref29} & \multicolumn{3}{c}{0.3069} & \multicolumn{3}{c}{8.05}    \\ \hline
				LFCC-LCNN \cite{ref29} & \multicolumn{3}{c}{0.3152} & \multicolumn{3}{c}{8.90}    \\ \hline
				MPIF-Res2Net \textbf{ours}   & \multicolumn{3}{c}{\textbf{0.2557}} & \multicolumn{3}{c}{\textbf{3.29}}    \\ \hline
			\end{tabular}
		\end{table}
		
		\subsubsection{Performance Comparison With Other Systems}
		
		The Table ~\ref{result2} shows the  results on ASVspoof 2021 LA dataset of different fusion systems. Although the fusion systems T23 \cite{ref22}, T20 \cite{ref23}, T35 \cite{ref22} and T19 \cite{ref22} outperform than our proposed MPIF-Res2Net, but their fusion ways are very complicated. Such as the T23, it is composed by 12 other systems trained separately, and got fused with finely adjusted weight assignment at the score stage. The method we proposed is based on a single system, which is less complicated compared to the fusion systems. 
		Table ~\ref{result1} shows the EER result of single systems, the best EER result of other systems is 8.05\%, the method we proposed has improved the performance by 59\% relative to the RawNet2\cite{ref29}system.

		\section{Conclusion}
		In this paper, we achieve accurate and useful information discrimination from two aspects. On the one hand, Specmix helps the model to focus on the location of key information in the sample by mixing information between samples, and randomly selects samples for Specmix operations, effectively avoiding the phenomenon of performance degradation caused by the destruction of original data. On the other hand, MPIF-Res2Net reduces redundant information caused by learning similar information from a single perspective by fusing information from multiple perspectives, removing the influence of redundant information on the learning of key information. The effectiveness of our method has been demonstrated by experiments. The effectiveness of our proposed method was verified by the experiment results. 
		


	\end{CJK*}
\end{document}